\documentclass[preprint,aps,showpacs,nofootinbib,tightenlines]{revtex4}
\usepackage{amsmath}
\usepackage{amssymb}
\usepackage{epsfig}
\usepackage{graphicx}
\begin{document}
%%%%%%%%%%%%%%%%%%%%%%%%%%%%%%%%%%%%%%%%%%%%%

\newcommand{\beq}{\begin{eqnarray}}
\newcommand{\eeq}{\end{eqnarray}}
\newcommand{\non}{\nonumber\\ }
\newcommand{\acp}{{\cal A}_{CP}}
\newcommand{\ov}{\overline}
\newcommand{\jpsi}{J/\psi}
\newcommand{\etap}{\eta^{(\prime)}}

\newcommand{\psl}{ P \hspace{-2.8truemm}/ }
\newcommand{\nsl}{ n \hspace{-2.2truemm}/ }
\newcommand{\vsl}{ v \hspace{-2.2truemm}/ }
\newcommand{\epsl}{\epsilon \hspace{-1.8truemm}/\,  }

\def \epjc{ Eur. Phys. J. C }
\def \jpg{  J. Phys. G }
\def \npb{  Nucl. Phys. B }
\def \npps { Nucl. Phys. B (Proc. Suppl.) }
\def \plb{  Phys. Lett. B }
\def \pr{  Phys. Rep. }
\def \prd{  Phys. Rev. D }
\def \prl{  Phys. Rev. Lett.  }
\def \zpc{  Z. Phys. C  }
\def \jhep{ J. High Energy Phys.  }
\def \ijmpa { Int. J. Mod. Phys. A }
\def \cpc{ Chin. Phys. C }
\def \ctp{ Commun. Theor. Phys. }
\def \rmp{ Rev. Mod. Phys. }
\def \ppnp{ Prog. Part. $\&$ Nucl. Phys. }
\def \arnps{ Ann. Rev. Nucl. Part. Sci. }

%%%%%%%%%%%%%%%%%%%%%%%%%%%%%%%%%%%%%%%%%%%%%%%%%%%%
%%
\title{ Implications on $\eta$-$\eta'$-glueball mixing from $B_{d/s} \to \jpsi \etap$
Decays }
\author{
Xin~Liu$^1$\footnote{Electronic address: liuxin.physics@gmail.com},
Hsiang-nan~Li$^{2,3,4}$\footnote{Electronic address: hnli@phys.sinica.edu.tw},
and
Zhen-Jun~Xiao$^5$\footnote{Electronic address: xiaozhenjun@njnu.edu.cn}}
\affiliation{
$^1$School of Physics and Electronic Engineering, Jiangsu Normal University,\\ Xuzhou, Jiangsu 221116,
People's Republic of China\\
$^2$Institute of Physics, Academia Sinica, Taipei, Taiwan 115, Republic of China\\
$^3$Department of Physics, Tsing-Hua University, Hsinchu, Taiwan 300, Republic of China\\
$^4$Department of Physics, National Cheng-Kung University, Tainan, Taiwan 701, Republic of China\\
$^5$Department of Physics and Institute of Theoretical
Physics, Nanjing Normal University, Nanjing, Jiangsu 210046,
People's Republic of China}
\date{\today}
\bigskip
\begin{abstract}
\bigskip

We point out that the recent Belle measurements of the $B_{d/s} \to
\jpsi \etap$ decays imply large pseudoscalar glueball contents in
the $\eta^{(\prime)}$ meson. These decays are studied in the
perturbative QCD (PQCD) approach, considering the $\eta$-$\eta'$-$G$
mixing, where $G$ represents the pseudoscalar glueball. It is shown
that the PQCD predictions for the $B_{d/s} \to \jpsi \etap$
branching ratios agree well with the data for the mixing angle
$\phi_G\approx 30^\circ$ between the flavor-singlet state and the
pure pseudoscalar glueball. Extending the formalism to the
$\eta$-$\eta'$-$G$-$\eta_c$ tetramixing, the abnormally large
observed $B_d\to K\eta'$ branching ratios are also explained. The
proposed mixing formalism is applicable to other heavy meson decays
into $\eta^{(\prime)}$ mesons, and could be tested by future LHCb
and Super-$B$ factory data.

\end{abstract}

\pacs{13.25.Hw, 12.38.Bx, 14.40.Nd}

\maketitle

Recently, the Belle Collaboration reported their new measurements of
the $B_d\to \jpsi \etap$ decays~\cite{Chang12:BelleExp}, and the
first observations of the $B_{s} \to \jpsi \etap$
decays~\cite{Li12:BelleExp} with the branching ratios \beq {\rm
BR}(B_d \to \jpsi \eta)_{\rm Exp} &=& 12.3^{+1.9}_{-1.8}
\;\times 10^{-6}\;,\label{eq:br-bd2jpsie-ex} \\
{\rm BR}(B_d \to \jpsi \eta')_{\rm Exp} &<&
\hspace{0.5cm}7.4\hspace{0.5cm} \times 10^{-6} \;,
\hspace{0.5cm}({\rm 90\%\;\; C. L.})\label{eq:br-bd2jpsiep-ex}
\eeq
\beq
{\rm BR}(B_s \to \jpsi \eta)_{\rm Exp} &=& 5.10\pm 0.50(stat.)
\pm 0.25(syst.)^{+1.14}_{-0.79}(N_{B_s^{(*)}\bar{B}_s^{(*)}})
\;\times 10^{-4}\;,\label{eq:br-bs2jpsie-ex} \\
{\rm BR}(B_s \to \jpsi \eta')_{\rm Exp} &=& 3.71\pm 0.61(stat.)
\pm 0.18(syst.)^{+0.83}_{-0.57}(N_{B_s^{(*)}\bar{B}_s^{(*)}})
\; \times 10^{-4}\;,\label{eq:br-bs2jpsiep-ex}
 \eeq
and the relation of the branching ratios~\cite{Li12:BelleExp} %%%
\beq
{\rm R_s^{Exp}} &=& \frac{{\rm BR}(B_s \to \jpsi \eta')}{{\rm
BR}(B_s \to \jpsi \eta)} = 0.73\pm 0.14 (stat.) \pm 0.02(syst.)\;.
\label{eq:ra-br-bspsi-ex}
 \eeq
The updated result in Eq.~(\ref{eq:br-bd2jpsie-ex}), as stated
in~\cite{Chang12:BelleExp}, is consistent with and supersedes the
previous one, BR$(B_d \to\jpsi \eta) = 9.5^{+1.9}_{-1.9} \times
10^{-6}$, in 2007~\cite{Chang07:BelleExp}. The accuracy of the above
data is expected to be improved rapidly along with operation of LHCb
experiments.

The $B_{d/s} \to \jpsi \etap$ decays were first investigated by
Deandrea {\it et al.}~\cite{Deandrea93:psietap} under the naive
factorization assumption without considering the $\eta$-$\eta'$
mixing. In Ref.~\cite{Skands01:psieta}, Skands analyzed the $B_{d/s}
\to \jpsi \eta$ decays via SU(3) relations to the "golden channel"
$B_d \to \jpsi K^0$ including the $\eta$-$\eta'$ mixing. The
predictions BR$(B_d \to \jpsi \eta) = 11.0 \times 10^{-6}$ and
BR$(B_s \to \jpsi \eta) = 5.0 \times 10^{-4}$ (BR$(B_d \to \jpsi
\eta) = 15.0 \times 10^{-6}$ and BR$(B_s \to \jpsi \eta) = 3.3
\times 10^{-4}$), corresponding to $\theta = -10^\circ$ ($\theta =
-20^\circ$), were obtained~\cite{Skands01:psieta}, where $\theta$ is
the angle between the flavor-singlet state $\eta_1$ and the
flavor-octet state $\eta_8$. Note that the $B_{d/s} \to \jpsi \eta'$
branching ratios were not calculated in~\cite{Skands01:psieta}. If
they were, the outcome of BR$(B_{s} \to \jpsi \eta')$ would be much
larger than in Eq.~(\ref{eq:br-bs2jpsiep-ex}). This must be the case
in the conventional $\eta$-$\eta'$ mixing formalism as elaborated
below.

%%%============================================================
\begin{figure}[h!]
\vspace{-0.25cm} \centerline{\epsfxsize=15 cm \epsffile{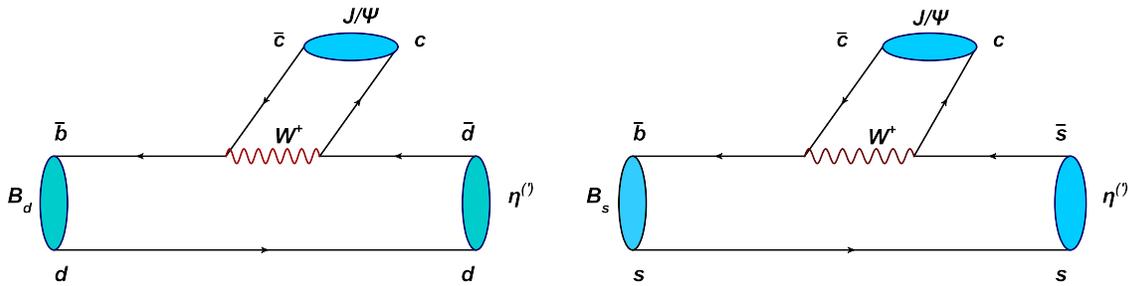}}
\vspace{0.1cm} \caption{(Color online) Leading quark-level
diagrams for the $B_{d/s} \to \jpsi
\etap$ decays.} \label{fig:fig1}
\end{figure}
%%%============================================================

The $\eta^{(\prime)}$ mesons are produced via their nonstrange
components in the $B_d$ meson decays at leading level, while they
are produced via their strange components in the $B_s$ meson decays,
as easily seen in Fig.~\ref{fig:fig1}. Under the conventional
mixing, the physical states $\eta$ and $\eta^\prime$ are related to
the quark-flavor states $\eta_q= (u \bar u + d \bar d)/\sqrt{2}$ and
$\eta_s= s \bar s$ through a mixing matrix with an angle
$\phi$~\cite{Feldmann98-99:etap}, \beq \left(
\begin{array}{c} \eta\\ \eta^\prime \\ \end{array} \right ) &=&
  \left( \begin{array}{cc}
 \cos{\phi} & -\sin{\phi} \\
 \sin{\phi} & \cos\phi \end{array} \right )
 \left( \begin{array}{c}  \eta_q\\ \eta_s \\ \end{array} \right )\;.
 \label{eq:mix-eta-etap}
 \eeq
Various experimental and theoretical constraints have yielded $\phi
< 45^\circ$ (for a recent detailed discussion, see
Ref.~\cite{Donato11:etap}), for which one observes
\cite{Datta02:psietap} \beq
 {\rm R_d} &\equiv& \frac{{\rm BR} (B_d \to \jpsi \eta')}{{\rm BR} (B_d \to \jpsi \eta)}
 \approx \tan^2 \phi <1\;, \label{eq:rd-psietap} \\
 {\rm R_s} &\equiv& \frac{{\rm BR} (B_s \to \jpsi \eta')}{{\rm BR} (B_s \to \jpsi \eta)}
 \approx \cot^2 \phi >1\;. \label{eq:rs-psietap}
  \eeq
It is obvious that only the first relation meets the data, and the
second one does not as confronted with
Eq.~(\ref{eq:ra-br-bspsi-ex}). As speculated in
\cite{Li12:BelleExp}, the small measured ratio ${\rm R_s^{Exp}}$
indicates additional flavor-singlet components in the $\eta'$ meson
other than the $u \bar u$, $d \bar d$ and $s \bar s$ pairs, or
violation of the $\eta$-$\eta'$ mixing scheme.

Inspired by the Belle measurements, we analyze the $B_{d/s} \to
\jpsi \etap$ decays in the more complete $\eta$-$\eta'$-$G$ mixing
formalism, $G$ denoting a pseudoscalar glueball. Adopt the
construction in~\cite{Li08:etap},
\begin{equation}
\label{qs}
   \left( \begin{array}{c}
    |\eta\rangle \\ |\eta'\rangle\\|G\rangle
   \end{array} \right)
   = U_3(\theta)U_1(\phi_G)
   \left( \begin{array}{c}
    |\eta_8\rangle \\ |\eta_1\rangle\\|g\rangle
   \end{array} \right) \;,
\end{equation}
where the matrices
\begin{equation}
U_1(\phi_G)=\left( \begin{array}{ccc}
    1 &0 &0\\
    0 &\cos\phi_G & \sin\phi_G \\
    0 &-\sin\phi_G & \cos\phi_G
   \end{array} \right)\;,\;\;\;\;
U_3(\theta)=\left( \begin{array}{ccc}
    \cos\theta & -\sin\theta & 0\\
    \sin\theta & \cos\theta &0\\
    0 &0&1
   \end{array} \right)\;,
\end{equation}
represent rotations around the axis along the $\eta_8$ meson and the
unmixed glueball $g$, respectively. Equation~(\ref{qs}) is based on
the assumption that $\eta_8$ does not mix with the glueball, under
which two mixing angles $\theta$ and $\phi_G$ are sufficient.

The octet and singlet states are related to the flavor states
through
\begin{equation}
\left( \begin{array}{c}
    |\eta_8\rangle \\ |\eta_1\rangle\\|g\rangle
   \end{array} \right)
   = U_3(\theta_i)
\left( \begin{array}{c}
    |\eta_q\rangle \\ |\eta_s\rangle\\|g\rangle
   \end{array} \right) \;,
\end{equation}
where $\theta_i$ is the ideal mixing angle with
$\cos\theta_i=\sqrt{1/3}$, i.e., $\theta_i=54.7^\circ$. The flavor
states are then transformed into the physical states via the matrix
\begin{eqnarray}
U(\phi,\phi_G)&=&U_3(\theta)U_1(\phi_G)U_3(\theta_i)\;,\nonumber\\
&=&\left(
\begin{array}{ccc}
\cos\phi+\sin\theta\sin\theta_i\Delta_G & -\sin\phi+\sin\theta\cos\theta_i\Delta_G & -\sin\theta\sin\phi_G\\
\sin\phi-\cos\theta\sin\theta_i\Delta_G & \cos\phi-\cos\theta\cos\theta_i\Delta_G & \cos\theta\sin\phi_G\\
  -\sin\theta_i\sin\phi_G &-\cos\theta_i\sin\phi_G&\cos\phi_G
   \end{array} \right)\;,\label{mut}
\end{eqnarray}
with the angle $\phi=\theta+\theta_i$ and the abbreviation
$\Delta_G\equiv 1-\cos\phi_G$. $U$ has been written in the form,
that approaches the mixing matrix in Eq.~(\ref{eq:mix-eta-etap}) in
the $\phi_G\to 0$ limit. We mention that a different mixing matrix
has been constructed by diagonalizing a mass matrix for pseudoscalar
bound states, which was derived from an effective QCD Lagrangian
\cite{MV10}.

It has been verified that the contribution from the gluonic
distribution amplitudes in the $\eta^{(\prime)}$ meson is negligible
for $B$ meson transition form factors \cite{Charng:2006zj}. Hence,
the $\eta$ and $\eta^{\prime}$ mesons are still produced via the
nonstrange (strange) component in the $B_d$ ($B_s$) meson decays
under the $\eta$-$\eta'$-$G$ mixing. Equation~(\ref{mut}) then
simply modifies Eqs.~(\ref{eq:rd-psietap}) and (\ref{eq:rs-psietap})
into
 \beq
 {\rm R_d^{Th}} &\approx&
\biggl(\frac{\sin\phi-\cos\theta\sin\theta_i\Delta_G}
{\cos\phi+\sin\theta\sin\theta_i\Delta_G}\biggr)^2
\;, \label{eq:rd}\\
 {\rm R_s^{Th}} &\approx&
\biggl(\frac{\cos\phi-\cos\theta\cos\theta_i\Delta_G}
{-\sin\phi+\sin\theta\cos\theta_i\Delta_G}\biggr)^2 \;,
\label{eq:rs}
 \eeq
respectively. Since the angle $\theta$ is negative, both the
numerator and the denominator in Eq.~(\ref{eq:rd}) decrease, and the
ratio ${\rm R_d^{Th}}$ could remain smaller than unity. On the other
hand, the numerator in Eq.~(\ref{eq:rs}) decreases, while the
magnitude of denominator increases. Therefore, the ratio ${\rm
R_s^{Th}}$ might drop from above unity to below unity for a
sufficiently large angle $\phi_G$.

We compute the $B_{d/s} \to \jpsi \etap$ decays explicitly in
the $\eta$-$\eta'$-$G$ mixing formalism, employing the perturbative
QCD (PQCD) approach~\cite{Keum01:kpi} at next-to-leading order (NLO) of
the strong coupling constant. Replacing the kinematic
variables and distribution amplitudes of the $K^0$ meson by those
of the $\eta_{q(s)}$ meson, we derive the $B_d \to \jpsi \etap$ decay amplitudes from
the $B_d \to \jpsi K^0$ ones~\cite{Liu10:psik}. Further substituting
$B_s$ for $B_d$, we arrive at the $B_{s} \to \jpsi \etap$
decay amplitudes. Throughout this work it is assumed that there are
no final state interactions and no isospin violation.
% (namely, with the equal $u$ and $d$ quark masses, $m_u = m_d$).
The input parameters such as the QCD scale~(in units of {\rm GeV}),
masses~({\rm GeV}), decay constants (GeV), and $B$ meson lifetime
({\rm ps}) are set to~\cite{Keum01:kpi,Nakamura10:pdg} \beq
 \Lambda_{\overline{\mathrm{MS}}}^{(f=4)} &=& 0.287\;,
 \qquad m_W = 80.41\;,  \qquad m_{B_d} = 5.28\;,
 \qquad m_{B_s} = 5.37\;;\non
   m_{\jpsi} &=& 3.097\;,\hspace{0.62cm} \quad m_b = 4.8 \;,
   \hspace{1.08cm} \quad   m_{c} = 1.50\;,
 \hspace{-0.02cm}  \qquad f_{\jpsi}= 0.405\;; \non
   f_{B_d} &=& 0.21\;,\hspace{0.76cm} \quad f_{B_s} = 0.23 \;,
   \hspace{0.85cm} \quad   \tau_{B_d} = 1.53\;,
 \hspace{0.20cm}  \qquad \tau_{B_s}= 1.47\;;\non
    f_{q} &=& f_\pi\;,\hspace{1.32cm} \quad f_s = 1.3 f_\pi \;,
   \hspace{0.55cm} \qquad   \phi = 43.7^\circ\;,
 \hspace{0.10cm}  \qquad \phi_G= 33^\circ\;; \non
   m_0^{\eta_q} &=& 1.50\;,\hspace{0.7cm} \quad m_0^{\eta_s} = 2.00 \;,
   \hspace{0.60cm} \qquad   f_\pi = 0.13\;. \label{eq:mass}
\eeq Following the observation in \cite{TLQ}, we have chosen a value
for $m_0^{\eta_q}$ slightly higher than the chiral scale for the
pion, $m_0^\pi\approx 1.3\sim 1.4$ GeV. The best fit on the mixing
angle $\theta$ from experiment data converged in the range of
$[-11^\circ, -17^\circ]$~\cite{Kou01:etap}. Our choice
$\phi=43.7^\circ$ in Eq.~(\ref{eq:mass}) is equivalent to
$\theta=\phi-\theta_i=-11^\circ$, close to $\theta=-10^\circ$
in~\cite{Skands01:psieta}, which corresponds to a sizable gluonic
admixture. It is also consistent with $\phi = (44^{+6}_{-7})^\circ$
determined from the $B_d \to \jpsi \eta/\pi^0$
data~\cite{Thomas07:etap}. The angle $\phi_G$, as summarized
in~\cite{Li08:etap}, varies in a wide range $10^\circ \lesssim
\phi_G \lesssim 30^\circ$. We have taken the central value of
$\phi_G = 33^\circ \pm 13^\circ$ recently extracted
in~\cite{Thomas07:etap,Escribano10:phig}. For the
Cabibbo-Kobayashi-Maskawa matrix elements, we employ the Wolfenstein
parametrization with the updated parameters $A=0.832$,
$\lambda=0.2246$, $\bar{\rho}=0.130$, and
$\bar{\eta}=0.350$~\cite{Nakamura10:pdg}.

The CP-averaged branching ratios for the $B_{d/s} \to \jpsi \etap$
decays in the standard model then read as,
  \beq
{\rm BR} (B_d \to \jpsi \eta) &=& 11.2^{+2.8}_{-2.1}
(\omega_{B_d})^{+1.9}_{-1.0}(a_2)^{+1.5}_{-1.4}(f_{\jpsi})
\times 10^{-6} \;, \label{eq:br-bd2psie-th} \\
{\rm BR} (B_d \to \jpsi \eta')&=&\hspace{0.24cm} 6.5^{+1.6}_{-1.2}
(\omega_{B_d})^{+1.1}_{-0.6}(a_2)^{+0.9}_{-0.8}(f_{\jpsi}) \times 10^{-6} \;, \label{eq:br-bd2psiep-th}\\
{\rm BR} (B_s \to \jpsi \eta) &=&  5.14^{+1.45}_{-1.10}
(\omega_{B_s})^{+1.10}_{-0.77}(a_2)^{+0.71}_{-0.64}(f_{\jpsi})
\times 10^{-4} \;, \label{eq:br-bs2psie-th} \\
{\rm BR} (B_s \to \jpsi \eta')&=&  3.68^{+1.04}_{-0.78}
(\omega_{B_s})^{+0.78}_{-0.55}(a_2)^{+0.51}_{-0.46}(f_{\jpsi}) \times
10^{-4} \;, \label{eq:br-bs2psiep-th}
  \eeq
with the ratios ${\rm R_d^{Th}}\approx 0.58$ and ${\rm
R_s^{Th}}\approx 0.72$. We have kept the $\eta'$ meson mass in the
phase-space factor for the $B_{d/s} \to \jpsi \eta'$ decay rates,
and neglected masses of other light mesons. The theoretical
uncertainties arise from the variation of the shape parameter
$\omega_B = 0.40 \pm 0.04$ $(\omega_{B_s}= 0.50 \pm 0.05)$ GeV for
the $B_d$ $(B_s)$ meson wave function \cite{LMS05,Ali07:pqcd}, of
the $\jpsi$ meson decay constant $f_{\jpsi}  = 0.405 \pm 0.014$ GeV
\cite{CL05}, and of the Gegenbauer coefficient $a_2 = 0.44 \pm 0.22$ 
for the leading-twist $\eta_{q(s)}$ distribution amplitude.
Obviously, the theoretical branching ratios are in good agreement
with the existing data and upper bound after considering the
$\eta$-$\eta'$-$G$ mixing.

A remark is in order. The $B_{s} \to \jpsi \etap$ decays have been
proposed~\cite{FKR11} to explore the $\eta$-$\eta'$-$G$ mixing with
KLOE's parametrization for the mixing matrix~\cite{KLOE}, namely,
assuming the absence of the glueball content in the $\eta$ meson. It
was observed that the angles $\phi\approx 40^\circ$ and $\phi
_G\approx 20^\circ$ determined by KLOE lead to ${\rm R_s}\approx 1$,
consistent with the previous Belle data \cite{Belle_old}. Hence, the
updated data ${\rm R_s}< 1$ indeed imply larger glueball components
in the $\etap$ meson as explained before. In principle, ${\rm R_s} <
1$ can be achieved in both KLOE and our parametrizations by tuning
the angles. Our parametrization is close to that
in~\cite{Thomas07:etap,Escribano10:phig} with a nonvanishing
glueball component in the $\eta$ meson, so we chose the larger
$\phi_G=33^\circ$ extracted
in~\cite{Thomas07:etap,Escribano10:phig}, and obtained ${\rm R_s} <
1$ naturally. Taking $\phi _G= 22^\circ$~\cite{KLOE}, the central
values of the branching ratios become
 \beq
 {\rm BR}(B_d \to \jpsi \eta) &=& 11.7 \times 10^{-6} \;, \quad
 {\rm BR}(B_d \to \jpsi \eta') = \hspace{0.2cm}8.2 \times 10^{-6}\;, \\
 {\rm BR}(B_s \to \jpsi \eta) &=& 5.00 \times 10^{-4}\;, \quad
 {\rm BR}(B_s \to \jpsi \eta') = 4.28 \times 10^{-4}\;,
 \eeq
and the consistency with the data deteriorates. Compared to
Eqs.~(\ref{eq:br-bd2psiep-th}) and (\ref{eq:br-bs2psiep-th}), we
find that BR$(B_{d/s} \to \jpsi \eta)$ are less sensitive to
$\phi_G$ than BR$(B_{d/s} \to \jpsi \eta')$, a result attributed to
the smaller glueball component in the $\eta$ meson. Another remark
is that as far as the $B_d \to \jpsi \etap$ decays are concerned,
their branching ratios can be accommodated in the conventional
$\eta$-$\eta'$ mixing by tuning the angle $\phi$. Fitting
Eq.~(\ref{eq:rd-psietap}) to our predictions in
Eqs.~(\ref{eq:br-bd2psie-th}) and (\ref{eq:br-bd2psiep-th}) yields
$\phi\approx 37.3^\circ$, close to $\phi=39.3^\circ \pm 1.0^\circ$
in~\cite{Feldmann98-99:etap}.

It is interesting to examine whether $D,D_s$ decays into $\etap$
mesons, such as $D,D_s \to \etap\ell^+\nu$, reveal the similar implication on the
mixing mechanism~\cite{Donato11:etap,CCD09,KLW10}. The ratios
of the branching ratios of these semileptonic decays have been
expressed as
\begin{eqnarray}
{\rm R_{d}^\prime}&\equiv& \frac{{\rm BR}(D^+ \to \eta'\ell^+\nu)}{{\rm BR} (D^+
\to \eta\ell^+\nu)}=\tilde R_D\tan^2\phi\;,\\
{\rm R_{s}^\prime}&\equiv& \frac{{\rm BR}(D_s \to \eta'\ell^+\nu)}{{\rm BR} (D_s
\to \eta\ell^+\nu)}=R_D\cot^2\phi\;,
\end{eqnarray}
where the factors $\tilde R_D\approx R_D$ collect the information
on the $D_{d/s}\to\eta_{q/s}$ transition form factors and the corresponding
phase space. The estimate of $R_D$ depends on how to
model the $q^2$ dependence of the form factor, $q^2$ being the lepton-pair
invariant mass squared, and suffers theoretical uncertainty.
Taking $R_D\approx 0.28$ \cite{ABM97}, it was found that the measured
values ${\rm R_{d}^{\prime Exp}}\approx 0.19 \pm 0.05 <R_D$ \cite{JY11} and
${\rm R_{s}^{\prime Exp}}=0.37\pm 0.10>R_D$~\cite{Nakamura10:pdg} exhibit
a pattern in agreement with the conventional $\eta$-$\eta'$ mixing.
However, we stress that the above observation is not in conflict with
the $\eta$-$\eta'$-$G$ mixing formalism advocated in this work, viewing
the potential uncertainties in the estimate of $R_D$ and in the assumption
$\tilde R_D \approx R_D$. In addition, it is not sure that the contributions
from the $D,D_s$ transitions to pseudoscalar glueballs, which were referred as
the weak annihilation process in~\cite{Donato11:etap}, are negligible as in
the $B_{d/s}$ meson decays. The inclusion of these channels would modify the analysis
of the $D,D_s \to \etap\ell^+\nu$ decays.

At last, we have
confirmed that the larger angle $\phi_G=33^\circ$ generates the
pseudoscalar glueball mass $m_G\approx 1.49$ GeV, following the
$\eta$-$\eta'$-$G$ mixing formalism in~\cite{Li08:etap}. Namely, the
postulation~\cite{Li08:etap} that the the $\eta(1405)$ meson is a
leading candidate for the pseudoscalar glueball is not altered. A
larger $\phi_G$ also makes an impact on the $B\to K\eta^{(\prime)}$
branching ratios in the $\eta$-$\eta'$-$G$-$\eta_c$ tetramixing
formalism~\cite{TLQ}. Substituting $\phi_G=33^\circ$ into the
formula in Sec.~III D of \cite{TLQ}, BR$(B^0\to K^0\eta^{\prime})$
is enhanced from $50\times 10^{-6}$ in NLO PQCD \cite{Xiao08:ketap}
to $(59.7^{+22.6}_{-16.4})\times 10^{-6}$. On the other hand,
BR$(B^0\to K^0\eta)$ is insensitive to the variation of $\phi_G$,
which remains as around $2\times 10^{-6}$~\cite{TLQ}. Therefore, the
dramatically different data for BR$(B^0\to
K^0\eta^{\prime})=(66.1\pm 3.1)\times 10^{-6}$ and BR$(B^0\to
K^0\eta)=(1.12^{+0.30}_{-0.28})\times 10^{-6}$~\cite{HFAG} can also
be understood.

In summary, most of studies in the literature on $B_{d/s}$ decays
into $\eta^{(\prime)}$ mesons were performed under the conventional
$\eta$-$\eta'$ mixing. The recent $B_{d/s} \to \jpsi \etap$ data
provided a strong implication on the sizable pseudoscalar glueball
contents in the $\eta^{(\prime)}$ meson, which motivated our
investigation in the $\eta$-$\eta'$-$G$ mixing formalism. We
have verified this
implication by computing explicitly the $B_{d/s} \to \jpsi \etap$
branching ratios in the NLO PQCD approach: the outcomes from a
large angle $\phi_G\approx 30^\circ$ were found to be well
consistent with the current measurements and upper bounds. The
abnormally large observed $B\to K\eta'$ branching ratios were also
accommodated in the $\eta$-$\eta'$-$G$-$\eta_c$ tetramixing
formalism with the same $\phi_G$. Our work suggests that complete
understanding of dynamics in $\etap$-involved processes demands the
$\eta$-$\eta'$-$G$ mixing scheme. The resultant predictions for
other $B_{d/s}\to\eta^{(\prime)}$ decays could be tested by future
data of LHCb and/or Super-$B$ factories.

\begin{acknowledgments}

We thank C.H. Chen, F. De Fazio, R. Fleischer,  V. Mathieu, and
G. Ricciardi for useful discussions.
This work is supported, in part, by National Science Council of
R.O.C. under Grant No. NSC 98-2112-M-001-015-MY3, by the National
Natural Science Foundation of China under Grant No.~10975074, and
No.~10735080, by a project funded by the Priority Academic Program
Development of Jiangsu Higher Education Institutions (PAPD), and by
the Research Fund of Jiangsu Normal University under Grant
No.~11XLR38.

\end{acknowledgments}

%%%%%%%%%%%%%%%%%%%%%%%%%%%%%%%%%%%%%%%%%%%%%%%%%%%%%%%%%%%%%%%%%%%%%%%%%%%%%%%%%%
%                                        Appendix
%%%%%%%%%%%%%%%%%%%%%%%%%%%%%%%%%%%%%%%%%%%%%%%%%%%%%%%%%%%%%%%%%%%%%%%%%%%%%%%%5

%\begin{appendix}

%\end{appendix}

%%%%%%%%%%%%%%%%%%%%%%%%%%%%%%%%%%%%%%%%%%%%%%%%%%%%%%%%%%%%%%%%%%%%%%%%%%%%%%%%%%%%%%%%%%%%%%5
%                                 reference
%%%%%%%%%%%%%%%%%%%%%%%%%%%%%%%%%%%%%%%%%%%%%%%%%%%%%%%%%%%%%%%%%%%%%%%%%%%%%%%%%%%%%%%%%%%%%%%%%

\end{document}